\documentstyle[11pt,mypasp,twoside,epsf]{article}
\def\edcomment#1{\iffalse\marginpar{\raggedright\sl#1\/}\else\relax\fi}
\reversemarginpar

\begin{document}


\title{Cosmological Uses of Gamma-Ray Bursts}

\author{
S.G. Djorgovski$^*$,
S.R. Kulkarni$^*$,
D.A. Frail$^\dagger$,
F.A. Harrison$^*$,
J.S. Bloom$^\ddag$,
E. Berger$^*$,
P.A. Price$^*$,
D. Fox$^*$,
A.M. Soderberg$^*$,
R. Sari$^*$,
~~~S. Yost$^*$,
A.A. Mahabal$^*$,
S.M. Castro$^*$,
R. Goodrich$^\star$, 
F. Chaffee$^\star$
}
\affil{$^*$ California Institute of Technology, Pasadena, CA 91125, USA}
\affil{$^\dagger$ National Radio Astronomy Observatory, Socorro, NM 87801, USA}
\affil{$^\ddag$ Harvard-Smithsonian Ctr.for Astrophysics, Cambridge,MA\thinspace02138,USA}
\affil{$^\star$ W.M. Keck Observatory, Kamuela, HI 96743, USA}

\begin{abstract}
Studies of the cosmic gamma-ray bursts (GRBs) and their host galaxies are 
starting to provide interesting or even unique new insights in observational
cosmology.  GRBs represent a new way of identifying a population of 
star-forming galaxies at cosmological redshifts.  GRB hosts are broadly
similar to the normal field galaxy populations at comparable redshifts
and magnitudes, and indicate at most a mild luminosity evolution out to
$z \sim 1.5 - 2$.  GRB optical afterglows seen in absorption provide a
powerful new probe of the ISM in dense, central regions of their host
galaxies, complementary to the traditional studies using QSO absorbers.
Some GRB hosts are heavily obscured, and provide a new
way to select a population of cosmological sub-mm sources, and a
novel constraint on the total obscured fraction of star formation over
the history of the universe.  Finally, detection of GRB afterglows at
$z > 6$ may provide a unique way to probe the primordial star formation,
massive IMF, early IGM, and chemical enrichment at the end of the cosmic
reionization era.
\end{abstract}

\section{Introduction: The Birth of the GRB Cosmology}

Ever since the establishment of their cosmological nature
(Metzger et al. 1997), and considering their increasingly probable
relation to massive star formation (e.g., Paczy\'nski 1998, Totani 1997,
etc.), GRBs promised to become new probes of cosmology and galaxy evolution.
As the numbers of GRB redshifts and detected hosts and afterglows grow,
it becomes possible to use GRBs in new, systematic studies in cosmology.
There are now (late 2002) plausible or certain host galaxies found for
all but 1 or 2 of the bursts with optical, radio, or x-ray afterglows
localised with arcsec precision, and over 30 redshifts measured for GRB
hosts and/or afterglows, ranging from 0.25 (or perhaps 0.0085?) to 4.5,
with the median $\langle z \rangle \approx 1.0$.  GRBs and their afterglows
should be readly detectable at large redshifts (Lamb \& Reichart 2000).

There are three basic ways of learning about the evolution of luminous matter
and gas in the universe.  First, a direct detection of sources (i.e., galaxies)
in emission, either in the UV/optical/NIR (the unobscured component), or
in the FIR/sub-mm/radio (the obscured component).  Second, the detection of
galaxies selected in absorption along the lines of sight to luminous background
sources, traditionally QSOs.  Third, diffuse extragalactic backgrounds, which
bypass all of the flux or surface brightness selection effects plaguing all
surveys of discrete sources found in emission, but at a price of losing the
redshift information, and the ability to discriminate between the luminosity
components powered by star formation and powered by AGN.  Studies of GRB hosts
and afterglows can contribute to all three of these methodological approaches,
bringing in new, independent constraints for models of galaxy evolution and
of the history of star formation in the universe.

In this review we focus on some of the cosmological aspects of GRBs and
their host galaxies.  Parts of the present text have also appeared in the
reviews by Hurley, Sari \& Djorgovski (2003), and Djorgovski et al. (2003).
Some of the general issues regarding GRB redshifts and host galaxies have been
reviewed previously, e.g., by Djorgovski et al. (2001b, 2002), and many others.

\section{GRB Hosts and Galaxy Evolution}

The median apparent magnitude of GRB hosts
is $R \approx 25$ mag, with tentative detections or upper limits
reaching down to $R \approx 29$ mag.  The few missing cases are at least
qualitatively consistent with being in the faint tail of the observed
distribution of host galaxy magnitudes.
Down to $R \sim 25$ mag, the observed distribution is consistent with deep
field galaxy counts (Brunner, Connolly, \& Szalay 1999),
but fainter than that, complex selection effects may be playing a role.
It can also be argued that the observed distribution should correspond
roughly to luminosity-weighted field galaxy counts.  The actual
distribution would depend on many observational selection and physical
(galaxy evolution) effects, and a full interpretation of the observed
distribution of GRB host galaxy magnitudes requires a careful modeling
(see, e.g., Krumholz, Thorsett, \& Harrison 1998; Hogg \& Fruchter 1999).
The observed visible light (restframe UV) traces an indeterminate mix of
recently formed stars and an older population, and cannot be unambiguously
interpreted in terms of either the total baryonic mass, or the
instantaneous SFR, and their relative importance is a function of redshift.

Spectroscopic measurements provide direct estimates of recent, massive
SFR in GRB hosts, from
the luminosity of star formation powered recombination lines, notably
the [O II] 3727 doublet, Ly$\alpha$, and Balmer lines (Kennicut 1998),
the UV continuum at $\lambda_{rest} = 1500$ or 2800 \AA\ 
(Madau, Pozzetti, \& Dickinson 1998).
All of these estimators are susceptible to the internal extinction and its
geometry, and have an intrinsic scatter of at least 30\%.
The observed $unobscured$ SFR's range from a few tenths to a few 
$M_\odot$ yr$^{-1}$.  Applying the reddening corrections derived from the
Balmer decrements of the hosts, or from the modeling of the broad-band colors
of the OTs (and further assuming that they are representative of the mean
extinction for the corresponding host galaxies) increases these numbers
typically by a factor of a few.  All this is entirely 
typical for the normal field galaxy population at comparable redshifts.
However, such measurements are completely insensitive to any fully obscured
SFR components, which could be considerably higher.

Equivalent widths of the [O II] 3727 doublet in GRB hosts, which may provide
a crude measure of the SFR per unit luminosity (and a worse measure of the
SFR per unit mass), are on average somewhat higher (Djorgovski et al. 2001b)
than those observed in magnitude-limited field galaxy samples at comparable
redshifts (Hogg et al. 1998).
One intriguing hint comes from the flux ratios of [Ne III] 3869 to
[O II] 3727 lines: they are on average a factor of 4 to 5 higher in GRB
hosts than in star forming galaxies at low redshifts (Djorgovski et al. 2001b). 
Strong [Ne III] 
requires photoionization by massive stars in hot H II regions, and may represent
indirect evidence linking GRBs with massive star formation.

The interpretation of the luminosities and observed star formation rates is
vastly complicated by the unknown amount and geometry of extinction.  
Both observational windows, the optical/NIR (rest-frame UV) and the sub-mm
(rest-frame FIR) suffer from some biases: the optical band is significantly
affected by dust obscuration, while the sub-mm and radio bands lack
sensitivity, and therefore uncover only the most prodigiously star-forming
galaxies.  As of late 2002, radio and/or sub-mm emission powered by
obscured star formation has been detected from 4 GRB hosts
(Berger, Kulkarni \& Frail 2001; Berger et al. 2002b; Frail et al. 2002).
The surveys to date are sensitive only
to the ultra-luminous ($L > 10^{12} L_\odot$) hosts, with SFR of several
hundred M$_\odot$ yr$^{-1}$, but suggest that about 20\% of GRB hosts are
ULIRGs.

Given the uncertainties of the geometry of optically thin and optically
thick dust, optical colors of GRB hosts cannot be used to make any
strong statements about their net star formation activity.  The broad-band
optical colors of GRB hosts are not distinguishable from those of normal
field galaxies at comparable magnitudes and redshifts
(Bloom, Djorgovski, \& Kulkarni 2001; Sokolov et al. 2001; 
Chary, Becklin, \& Armus 2002; Le Floc'h et al. 2003).
It is notable that the optical/NIR colors of GRB hosts detected in the 
sub-mm are much bluer than typical sub-mm selected galaxies:
the GRB selection may be revealing a previously unprobed population of
dusty star-forming galaxies.

The magnitude and redshift distributions of GRB host galaxies are typical
for the normal, faint field galaxies (e.g., Schaefer 2000), as are their
morphologies (e.g., Odewahn et al. 1998, 
Bloom, Kulkarni \& Djorgovski 2002): they are often compact, and sometimes
suggestive of a merger (Djorgovski, Bloom \& Kulkarni 2001,
Hjorth et al. 2002), but that is not unusual for galaxies at comparable
redshifts.  The observed redshift distribution of GRBs is also at
least qualitatively as expected for an evolving, normal field galaxy
population (Mao \& Mo 1998).  

\begin{figure}
\null
\vskip 1.8 truecm
\plotfiddle{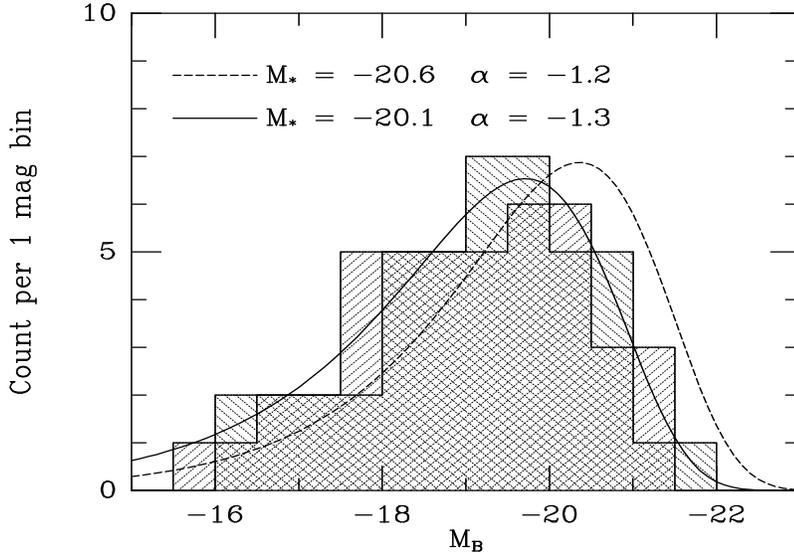}{5.0cm}{0}{60}{50}{-180}{-170}
\font\tenrm=cmr10
\caption{\tenrm
The observed distribution of GRB host galaxy absolute magnitudes in
the restframe $B$ band (histograms), fitted by the luminosity-weighted
Schechter luminosity function (solid line).  The two histograms represent
the same data, with the binning offset by half a bin.  The dashed line
corresponds to the  $z \sim 0$ GLF for all morphological types, from the
2df redshift survey.
}
\end{figure}

If GRB's follow the luminous mass, then the expected distribution would
be approximated by the luminosity-weighted galaxy luminosity
function (GLF) for the appropriate redshifts.
The hosts span a wide range of luminosities, with a characteristic absolute
restframe B band magnitude $M_{B,*} \approx -20$ mag, approximately half a
magnitude fainter than in the GLF at $z \approx 0$,
but comensurate with the late-type (i.e., star forming disk) galaxy
population at $z \approx 0$ (Madgwick et al. 2002; Norberg et al. 2002).  
This is somewhat surprising, since one
expects that the evolutionary effects would make the GRB host  galaxies,
with a typical $z \sim 1$, brighter than their descendants today.  The GRB
host GLF also has a somewhat steeper tail than the composite GLF at 
$z \approx 0$, but again similar to that of the star-forming, late-type
galaxies.  This is in a broad agreement with the results of deep redshift
surveys which probe the evolution of field galaxy populations out to $z \sim 1$
(Lilly et al. 1995; Fried et al. 2001; Lin et al. 1999), although our 
understanding of the field galaxy evolution in the same redshift range
is still very incomplete.  While much
remains to be done, it seems that GRB hosts provide a new, independent
check on the traditional studies of galaxy evolution at moderate and high
redshifts.

\section{GRBs as Probes of the Obscured Star Formation}

Already within months of the first detections of GRB afterglows, no OT's were
found associated with some well-localised bursts despite deep and rapid
searches; the prototype ``dark burst'' was GRB 970828 (Djorgovski et al. 2001a).
One explanation is that at least some of these ``missing'' afterglows are
obscured by dust in their host galaxies, which is certainly plausible if
GRBs are associated with massive star formation.  This is supported by
detections of RTs without OTs (e.g., Frail et al. 2000, 
Taylor et al. 2000).  Dust reddening has been detected directly in some OTs 
(Ramaprakash et al. 1998, Bloom et al. 1998, Djorgovski et al. 1998, etc.);
however, this only covers OTs seen through optically thin dust, and there
must be others, hidden by optically thick dust.
An especially dramatic case was the RT (Taylor et al. 1998) and 
IR transient (Larkin et al. 1998)
associated with GRB 980329 (Yost et al. 2002).
We thus know that at least some GRB OTs must be obscured by dust. 

This offers a possibility of making a
completely new and independent estimate of the mean obscured star formation
fraction in the universe.  The redshift distribution is
not a critical factor here; GRBs are now detected out to 
$z \sim 4.5$ and that there is no correlation of the observed fluence with 
the redshift (Djorgovski et al. 2002),
so GRBs are, at least to a first approximation, good probes of the star
formation over the observable universe.  

As of late 2002, there have been $\sim 70$ adequately deep and rapid
searches for OTs from well-localised GRBs, 
reaching at least to $R \sim 20$ mag within
less than a day from the burst, and/or to at least to $R \sim 23 - 24$ mag
within 2 or 3 days. 
In just over a half of such searches, OTs were found.
Inevitably, some OTs may have been missed due to an intrinsically low flux,
an unusually rapid decline rate (Fynbo et al. 2001; Berger et al. 2002a), 
or very high redshifts (so that the brightness in the commonly used $BVR$ bands
would be affected by the intergalactic absorption).
Thus the $maximum$ fraction of all OTs (and therefore massive star formation)
hidden by the dust is $\sim 50$\%.

This is a remarkable result.  It broadly agrees with the estimates that there
is roughly an equal amount of energy in the diffuse optical and FIR backgrounds
(see, e.g., Madau 1999).  This is contrary to some claims in the literature
which suggest that the fraction of the obscured star formation was much higher
at high redshifts.  Recall also that the fractions of the obscured and
unobscured star formation in the local universe are comparable.  

There is one possible loophole in this argument: GRBs may be able to destroy
the dust in their immediate vicinity (up to $\sim 10$ pc?)
(Waxman \& Draine 2000; Galama \& Wijers 2000),
and if the rest of the optical path through their hosts ($\sim$ kpc scale?)
was dust-free, OTs would become visible.  Such a geometrical arrangement may
be unlikely in most cases, and our argument probably still applies.
A more careful treatment of the dust evaporation geometry is needed, but
it is probably safe to say that GRBs can provide a valuable new constraint
on the history of star formation in the universe. 

\section{GRBs as Probes of the ISM in Evolving Galaxies}

Absorption spectroscopy of GRB afterglows is now becoming a powerful new
probe of the ISM in evolving galaxies, complementary to the traditional studies
of QSO absorption line systems.  The key point is that the GRBs almost by
definition (that is, if they are closely related to the sites of
ongoing or recent massive star formation, as the data seem to indicate)
probe the lines of sight to dense, central regions of their host galaxies
($\sim 1 - 10$ kpc scale).  On the other hand, the QSO absorption systems
are selected by the gas cross section, and favor large impact parameters
($\sim 10 - 100$ kpc scale), mostly probing the gaseous halos of field galaxies,
where the physical conditions are very different.

The associated GRB absorption systems show exceptionally high
column densities of gas, when compared to the typical QSO absorption systems;
only the highest column density DLA systems come close
(Savaglio, Fall \& Fiore 2002, Castro et al. 2003, Mirabal et al. 2002).
Lower redshift, intervening absorbers are also frequently seen, and their
properties appear to be no different from those of the QSO absorbers.
This opens the interesting prospect of using GRB absorbers as a new probe of
the chemical enrichment history in galaxies in a more direct fashion than
what is possible with the QSO absorbers, where there may be a very complex
dynamics of gas ejection, infall, and mixing at play.  

Properties of the GRB absorbers are presumably, but not necessarily (depending
on the unknown geometry of the gas along the line of sight) reflecting the ISM
of the circum-burst region.  Studies of their chemical composition do not yet
reveal any clear anomalies, or the degree of depletion of the dust, but the
samples in hand are still too small to be really conclusive.  Also, there
have been a few searches for the variability of the column density of the gas
on scales of hours to days after the burst, with no clear detections so far.
Such an effect may be expected if the burst afterglow modifies the physical
state of the gas and dust along the line of sight by the evaporation of the
dust grains, additional photoionization of the gas, etc.  However, it is
possible that all such changes are observable only on very short time scales,
seconds to minutes after the burst.  A clear detection of a
variable absorption against a GRB afterglow would be an important
result, providing new insight into the circumstances of GRB origins.

\section{High-Redshift GRBs:
A Unique Probe of the Primordial Star Formation and Reionization}

Possibly the most interesting use of GRBs in cosmology is as probes of the
early phases of star and galaxy formation, and the resulting reionization of
the universe at $z \sim 6 - 20$.  If GRBs reflect deaths of massive stars,
their very existence and statistics would provide a superb probe of the
primordial massive star formation and the initial mass function (IMF).  
They would be by far the most 
luminous sources in existence at such redshifts (much brighter than SNe, and
most AGN), and they may exist at redshifts where there were $no$ luminous
AGN.  As such, they would provide unique new insights into the physics and
evolution of the primordial IGM during the reionization era (see, e.g.,
Lamb \& Reichart 2001; Loeb 2002a,b).

There are two lines of argument in support of the existence of copious numbers
of GRBs at $z > 5$ or even 10.  First, a number of studies using photometric
redshift indicators for GRBs suggests that a substantial fraction (ranging
from $\sim 10$\% to $\sim 50$\%) of all bursts detectable by past,
current, or forthcoming missions may be originating at such high redshifts,
even after folding in the appropriate spacecraft/instrument selection
functions (Fenimore \& Ramirez-Ruiz 2002; Reichart et al. 2001;
Lloyd-Ronning, Fryer, \& Ramirez-Ruiz 2002).

Second, a number of modern theoretical studies suggest that the very first
generation of stars, formed through hydrogen cooling alone, were very
massive, with $M \sim 100 - 1000 ~M_\odot$ (Bromm, Coppi \& Larson 1999;
Abel, Bryan, \& Norman 2000; Bromm, Kudritzki, \& Loeb 2001;
Bromm, Coppi \& Larson 2002; Abel, Bryan \& Norman 2002).
While it is not yet absolutely clear that some as-yet unforseen effect would
lead to a substantial fragmentation of a protostellar object of such a mass, a
top-heavy primordial IMF is at least plausible.  It is also not yet 
completely clear that the (probably spectacular) end of such an object
would generate a GRB, but that too is at least plausible
(Fryer, Woosley \& Heger 2001).  Thus, there is some real hope that significant
numbers of GRBs and their afterglows would be detectable in the redshift range
$z \sim 5 - 20$, spanning the era of the first star formation and cosmic
reionization (Bromm \& Loeb 2002).

Spectroscopy of GRB aftergows at such redshifts would provide a crucial,
unique information about the physical state and evolution of the primordial
ISM during the reionization era.  The end stages of the cosmic reionization
have been detected by spectroscopy of QSOs at $z \sim 6$
(Djorgovski et al. 2001c; Fan et al. 2001; Becker et al. 2001).
GRBs are more useful in this context than the QSOs, for several reasons.
First, they may exist at high redshifts where there were no comparably
luminous AGN yet.  Second, their spectra are highly predictable power-laws,
without complications caused by the broad Ly$\alpha$ lines of QSOs, and can
reliably be extrapolated blueward of the Ly$\alpha$ line.  Finally, they would
provide a genuine snapshot of the intervening ISM, without an appreciable
proximity effect which would inevitably complicate the interpretation of
any high-$z$ QSO spectrum: luminous QSOs excavate their Stromgren spheres
in the surrounding neutral ISM out to radii of at least a few Mpc, whereas
the primordial GRB hosts would have a negligible effect of that type.

\acknowledgments

We wish to thank numerous collaborators, and the staff of Palomar and
W.M. Keck Observatories
for their expert help during our observing runs.
Our work was supported by grants from the NSF, NASA, and private donors.

\end{document}